\documentclass[pra,twocolumn,showpacs,amsmath,floatfix]{revtex4-1}
\usepackage{graphicx}%
\usepackage{dcolumn}
\begin{document}
\preprint{}
\title[Short Title ]{Zhu-Nakamura theory and the superparabolic level-glancing models\\
}

\author{J. Lehto}
\email{jaakko.lehto@utu.fi}
\affiliation{%
Turku Centre for Quantum Physics, Department of Physics and Astronomy, University of Turku,\\
FI-20014 Turku, Finland \\
}%

\date{\today}
\begin{abstract}
We study the applicability of the Zhu-Nakamura theory to a class of time-dependent quantum mechanical level-crossing models called superparabolic level-glancing models. The phenomenon of a level glancing, being on the borderline between a proper crossing of energy levels and an avoided crossing, is also an important special case between the two different approximative expressions in the Zhu-Nakamura theory. It is seen that the application of the theory to these models is not straightforward. We discuss some possible causes of these difficulties and also compare the approximative formulas of Zhu-Nakamura theory to those obtained by the generalization of the DDP theory.

\end{abstract}

\maketitle




\section{Introduction}
The level-crossing models form a paradigm in the study of quantum dynamics of nonadiabatic transitions \cite{NakamuraBook}. These models describe quantum systems with coupled states for which the corresponding diabatic energy levels, i.e, the energies related to the system eigenstates when there is no coupling present, depend on some external parameter and cross so that the states in question become degenerate at some parameter values. This type of situations appear both in time-independent form, for example, in the collisional problems where the nonadiabatic transitions happen effectively at the curve crossings of potentials that depend on a spatial coordinate \cite{Child1974}, as well as in purely time-dependent problems where the level crossings are induced by external time-dependent fields \cite{kasphd}. The proper frameworks for describing the dynamics in such situations are then the stationary and time-dependent Schr{\"o}dinger equations, respectively. 

The pioneering works on the subject were done by Landau, Zener, St{\"u}ckelberg and Majorana already in 1932 in their studies of the linear crossing problem in which the diabatic energies depend linearly on the external coordinate and the coupling between the diabatic states is constant \cite{Landau1932,Zener1932,Stuckelberg1932,Majorana1932}. This basic model is nowadays called the Landau-Zener (LZ) model. The relevance of this model, besides its relative simplicity, comes from the fact that the transitions are localized in the vicinity of the crossings so the above-mentioned behavior of the energy levels and the coupling often form a good approximation to describe the dynamics of many real physical systems. 
This and other level-crossing models have been widely applied over the years, for example, to the studies of atomic and molecular collisions \cite{Child1974, NakamuraBook}, laser-atom interactions \cite{kazantsev}, quantum information processing \cite{Gaitan2003} and in attempts to understand the dynamics of quantum phase transitions \cite{Dziarmaga2010, Polkovnikov2011}. 

One of the recent advances in the field of nonadiabatic transitions is the Zhu-Nakamura theory (ZNT) formulated by Nakamura and co-workers, which the authors claim to be a generally applicable and accurate approximate theory for any level-crossing model and, in this sense, to form a complete solution of the problem \cite{NakamuraBook, ZNTeranishi2001}. It is based on exact results obtained by Zhu and Nakamura for the linear curve-crossing problem, i.e., the time-independent LZ model \cite{ZNStokesConstants}. By generalizing the coupled wave integral method introduced by Hinton \cite{Hinton1979}, they were able to calculate the Stokes constants and thus to construct the scattering matrix for the problem. 
Unfortunately, the exact analytic expressions for the constants are very complicated and therefore not very useful. Because of this, they went on, starting from these exact results and working initially in the collisional setting, to build an approximate theory, i.e. ZNT, by introducing several phenomenological corrections to the final analytic formulas. 

In the recent years, the purely time-dependent problems especially have gained a lot of importance because of the progress of experimental methods and laser technologies in particular. This has lead to possibilities of controlling the state of a quantum system accurately, for example, by specifically tailored chirped laser pulses \cite{Zou2005, Chang2010}. Therefore, it is natural that the ZNT has been generalized to handle also these time-dependent problems \cite{tdznt}. This is due to the fact that the time-independent linear crossing model (i.e. LZ model) and the time-dependent quadratic crossing model (i.e. parabolic model \cite{Shimshoni1991, Suominen1992, Crothers1977}) both have the same analytical structure, namely, they both can be reduced to a triconfluent Heun equation \cite{Ronveaux}. This allows the same approximate formulas to be used in both settings. 
Despite the many different developments that are based on ZNT \cite{ZNNobusada2001, Nakamura2008} and the claim that the theory comprises a complete solution to curve-crossing problems,  there seems to be very few articles in the literature studying the basic characteristics of the theory. Also, the applicability of the theory must be in principle considered for each model independently due to the phenomenological character of the theory. 

In this paper, we consider a simple class of time-dependent level-crossing models, namely, the superparabolic level-glancing models introduced in \cite{artikkeli}. The parabolic time-dependencies of the diabatic energies have been applied e.g. in studies of laser-induced molecular dynamics \cite{Zou2005,Chang2010} and the parabolic level-glancing model was also recently used to study the tunneling between different energy bands in the case of merging Dirac cones \cite{Fuchs2012}. The level-glancing phenomenon can also be seen as a counterpart to the case where the energy matches exactly the crossing point energy in the time-independent models, a notoriously hard problem to approximate \cite{NakamuraBook, NikitinReview}. This particular parameter value remains as an important special case also in the ZNT, being the dividing parameter value between the two approximative formulas in the theory. While both of the formulas should be applicable to this case in principle, their connection is not entirely smooth (see, for example, Fig. 1 in Ref.~\cite{zn101}) and seems to be even less studied. 

We apply ZNT to superparabolic models and discuss particular aspects of the theory in this case. We also compare it to the other well-established approximative theory in the field of nonadiabatic transitions, namely, to the Dykhne-Davis-Pechukas theory \cite{dykhne, dp} studied in detail recently in \cite{artikkeli}. 
The structure of this paper is as follows. In the section \ref{sec:BasicEqs} we introduce the basic formalism for both the time-dependent and time-independent level-crossing problems and introduce the superparabolic level-glancing models. In the section \ref{sec:ddp}, the DDP theory and its application to superparabolic models is discussed shortly, more details can be found in \cite{artikkeli}. In the section \ref{sec:znt} we give a very condensed overview of the  ZNT and the final recommended approximate formulas in the form given in the basic reference \cite{NakamuraBook}. In section \ref{sec:results} we present and analyze the results that were obtained by numerical calculations and compare these to the approximative expressions. Finally, the discussion in Sec.~\ref{sec:conclusions} ends the presentation.

\section{Basic models and formalism}
\label{sec:BasicEqs}
\subsection{Time-Independent Two-State Processes}
\label{subsec:time-independent}

Although we concentrate on studying the time-dependent models in this paper, it is important to present also some of the main aspects of the time-independent LZ model in order to understand ZNT better. Here we give only a very minimal overview. For the details, the reader is directed to references.
 
The model describes a quantum system consisting of two states which both experience a potential that is linear in the coordinate $R$ and that have a constant coupling between them. That is, the system is governed by the time-independent Schr{\"o}dinger equation
\begin{equation}
-\frac{\hbar^{2}}{2m} \dfrac{d^{2}\varphi_{1}}{dR^{2}} + \left[ -F_{1}(R - R_{X}) - (E - E_{X})\right] \varphi_{1} = V_{12} \varphi_{2}
\label{eqn:scheq1}
\end{equation}
\begin{equation}
-\frac{\hbar^{2}}{2m} \dfrac{d^{2}\varphi_{2}}{dR^{2}} + \left[ -F_{2}(R - R_{X}) - (E - E_{X})\right] \varphi_{2} = V_{12} \varphi_{1},
\label{eqn:scheq2}
\end{equation}
with $F_{1} > 0 $, $V_{12} > 0$ and $F_{1} > F_{2}$. $F_{i}$'s are the slopes of the potentials, $V_{12}$ the diabatic coupling and $R_{X}$ and $E_{X}$ are the crossing point and the energy at the crossing point, respectively. The case where $F_{1}F_{2} > 0$ is called Landau-Zener type  and $F_{1}F_{2} < 0$ is called nonadiabatic tunneling type. We consider here only the LZ case because only it has a direct counterpart in the time-dependent theory. 

By transforming the coupled Eqs.~(\ref{eqn:scheq1}) and (\ref{eqn:scheq2}) into a momentum representation and redefining the variables suitably \cite{Child1974, NakamuraBook} 
we can reduce these coupled first-order equations to a second-order differential equation for, say, the state corresponding to $\varphi_{1}$, to a equation of the form
\begin{equation}
B''(z) + q(z)B(z) = 0,
\label{eqn:basicODE}
\end{equation}
where
\begin{equation}
q(z) = \frac{1}{4} - \dot{\imath} a^{2}z + \frac{1}{4}(a^{2} z^{2} - b^{2})^{2},
\label{eqn:coefficientfunction}
\end{equation}
This differential equation belongs to the class of triconfluent Heun equation \cite{Ronveaux}. The connection with the parameters in the original equations is obtained as
\begin{equation}
a^{2} = \frac{\hbar^{2}}{2 m}\frac{F\left(F_{1} - F_{2}\right)}{\left(2 V_{12}\right)^{2}}, \qquad b^{2} = \frac{\left(F_{1} - F_{2}\right)}{2FV_{12}}\left(E -E_{X} \right)
\label{eqn:param_ab}
\end{equation}
where it is defined $F = \sqrt{F_{1} \vert F_{2} \vert}$ and the independent variable is $ z = \left( 2 V_{12} k\right)/F$ where $k$ is the momentum. 

The important point here is only the physical meaning of these reduced parameters. The parameter $a^{2}$ represents the effective coupling strength and $b^{2}$ is the effective collision energy. From Eq.~(\ref{eqn:param_ab}) one can see that $a^{2}$ is always non-negative but that $b^{2}$ can be both positive or negative, depending on whether the energy $E$ is higher or lower than the crossing point energy $E_{X}$, respectively. In these variables, the celebrated LZ formula for the transition probability reads \cite{Zener1932}
\begin{equation}
p_{LZ} = \exp\left[-\frac{\pi}{4a\vert b\vert} \right].
\label{eqn:lz_formula}
\end{equation}
Zener obtained this result by reducing the time-independent problem described above to purely time-dependent one by approximating that the relative nuclear motion follows a straight-line trajectory with a constant velocity, i.e. $R(t) = v t$ where $v^{2} = 2\left(E-E_{X} \right)/m$. As is well known, this formula works only for energies much larger than the crossing point energy $E_{X}$. The physical reason is, that in a collision process the transition point is traversed twice and when $E$ approaches $E_{X}$ from above, the two transitions start to overlap. 

\subsection{Time-Dependent Two-State Processes}
\label{subsec:tdformalism}

Often considered canonical examples for purely time-dependent quantum mechanical problems are a spin$-\frac{1}{2}$ particle in a time-dependent magnetic field or two electronic states of an atom in a chirped or pulsed laser field. Another example is the collisional situation, explained in the previous subsection, with a heavy enough scattering particle so that it can be assigned with a distinct classical trajectory. 

In any case, whatever the underlaying physical situation is, the coherent dynamics of the two-state system is then given by the time-dependent Schr{\"o}dinger equation of the form: 
\begin{equation}
\dot{\imath} \hbar \dfrac{d}{dt}\varphi(t) = \begin{pmatrix} \varepsilon(t) & V(t) \\
					   V(t)  & -\varepsilon(t)\end{pmatrix} \varphi(t),
\label{eqn:scheq}
\end{equation}
and $\varphi(t) = \left[ c_{1}(t), c_{2}(t) \right]^{T} $, where $c_{1}(t)$ and $c_{2}(t)$ are the probability amplitudes of the diabatic basis states $\tilde{\varphi}_{1}$ and $\tilde{\varphi}_{2}$, respectively. The functions $\varepsilon(t)$ and $V(t)$ are similarly called as the diabatic energy levels and the diabatic coupling in this time-dependent setting. The crossings happen at points of time $t_{c}$ where $\varepsilon(t_{c}) = 0$. 

The Schr{\"o}dinger equation can be also given in the basis of the instantaneous eigenstates of the Hamiltonian matrix in (\ref{eqn:scheq}). In this adiabatic basis of the system, its energy levels, the adiabatic levels, are given as 
\begin{equation}
\mathcal{E}_{\pm} (t) = \pm \sqrt{\varepsilon(t)^{2} + V^{2}(t)},
\label{eqn:adiabaticlevels}
\end{equation}
and the adiabatic coupling reads
\begin{equation}
\gamma (t) = \pm \frac{ V(t) \dot{\varepsilon}(t) - \varepsilon(t) \dot{V}(t) }{2 \left( \varepsilon(t)^{2} + V(t)^{2}\right)},
\label{eqn:nonadiabcoupl}
\end{equation}
where the overhead dot stands for time derivative and one can fix the sign by fixing the relative sign of the basis vectors. In general, the functions $\varepsilon(t)$ and $V(t)$ do not have the same zeros, so the level crossing usually appears only as an avoided crossing in the adiabatic basis. Moreover, when we have 
\begin{equation}
\vert V(t) \vert \ll \vert \varepsilon (t)  \vert, \quad \vert t \vert \rightarrow \pm \infty, 
\end{equation}
the basis vectors of the two bases coincide, apart from the possible swap between the labels, and the initial and final probability distributions can be obtained from the same expression. In the superparabolic models studied in this paper, such a swapping of labels do not occur, and in the rest of this paper we take the initial conditions to be $\vert c_{2}\left( -\infty\right)\vert^{2} = 1$ so that both the diabatic and adiabatic transition probability, are given by $P \equiv \vert c_{1}\left( +\infty\right)\vert^{2}$.
\subsection{Superparabolic level-glancing models}

The simplest time-dependent model that can take into account the double-crossing process in the time-independent LZ model as explained previously, is the so-called parabolic model for which
\begin{equation}
\varepsilon(t) = \frac{At^{2} - B}{2}, \qquad V(t)= V_{0}.
\label{eqn:parabolicmodel}
\end{equation}
The parameters $A$, $V_{0}$ are positive while $B$ can also be negative. When  this happens, the diabatic energy levels do not cross and the transitions are possible only by tunneling while for positive $B$ the energy levels cross twice. In the limiting case between the two, namely when $B = 0$, the levels only touch each other at $t = 0$ and we call this a level-glancing case. 

Obviously, each of these cases are in a direct correspondence with the different cases in the time-independent LZ model. In fact, by making the following correspondencies between the variables and the parameters of these time-dependent and time-independent models 
\begin{equation}
t = z, \quad A = a^{2} \quad \textrm{and}\quad B = b^{2},
\label{eqn:corresp}
\end{equation}
and choosing the units in a such way that $\hbar = 1$ and $V_{0} = 1/2$, we see that differential equation for the probability amplitude $c_{2}$ obtained from Eq.~(\ref{eqn:scheq}) with (\ref{eqn:parabolicmodel}) is reduced to an equation that is completely equivalent form as (\ref{eqn:basicODE}). The same holds for $c_{1}$ but with the replacement $t = -z$. 
%

The parabolic model introduced above was the starting point for the ZNT, so it is expected that its approximate formulas work well in this case for all the parameter regions. Here we concentrate on the phenomenon of level glancing and set $B = 0$. Furthermore, we consider a direct generalization of the parabolic model, namely, the superparabolic models, where the diabatic energies are proportional to some even power of time. 
We can also reduce the number of parameters by one, and choose to work in units, where $\hbar = 1$ and the superparabolic level-glancing models are defined as \cite{artikkeli}
\begin{equation}
\varepsilon(t) = t^{N}, \qquad V(t)= \alpha = const,
\label{eqn:SPLGcmodel}
\end{equation}
where $N = 2, 4, 6, \ldots$ and $\alpha$ is positive. Now, the limit $\alpha \rightarrow 0$ is the sudden or diabatic limit while for large $\alpha$ the process is adiabatic. As explained at the end of the subsection (\ref{subsec:tdformalism}), in both these cases we have $P \approx 0$.

It should be noted that the definition of $\alpha$ is here different from the one used in \cite{NakamuraBook} or \cite{tdznt} for the parabolic model. We denote their parameter as $\alpha_{ZN} $ and it is actually equal to $A$ here. The connection between the two parameters is given by the relation
\begin{equation}
\alpha^{3} = 1/(4A).
\label{eqn:relparam}
\end{equation}
This difference in notation is of course unfortunate but we believe it is better to use the same notation as in \cite{artikkeli} for better exposition of the new results and for comparison between the existing ones. Furthermore, this should not give rise to additional confusion as the approximate formulas of ZNT for the time-dependent case have to be transformed from the expressions for the time-independent results by replacement of parameters anyway.
\begin{figure}
\begin{center} 
 \includegraphics[scale=0.72]{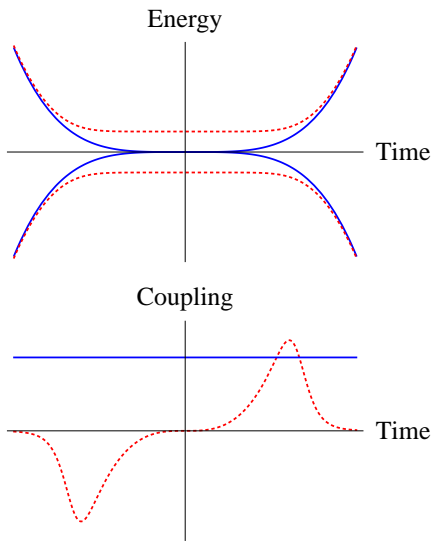} 
\caption{(Color online) Schematics of the time-dependence of the energy levels and coupling for the superparabolic models. The diabatic levels and the corresponding coupling are drawn with solid blue lines while the adiabatic ones are drawn with dashed lines in red.}
\label{fig:levelschematics}
\end{center}
\end{figure}

\section{Dykhne-Davis-Pechukas theory}
\label{sec:ddp}

One of the most important results concerning nonadiabatic transitions, and existing prior to ZNT, is given by a formula first proposed by Dykhne \cite{dykhne}, which connects the structure of the zeros $t_{c}$ of the adiabatic energies in the complex plane and the way the adiabatic limit is approached by the adiabatic transition probability. This was later proved rigorously for a class of two-level Hamiltonians by Davis and Pechukas \cite{dp}. The Dykhne-Davis-Pechukas (DDP) formula is  given by
\begin{equation}
P = e^{- 2 Im D(t_{c})},
\label{eqn:dykhne}
\end{equation}
where
\begin{equation}
D(t) =  \int_{0}^{t} \left( \mathcal{E}_{+}(s) - \mathcal{E}_{-}(s) \right) ds,
\label{eqn:D}
\end{equation}
and $t_{c}$ is defined by the equation
\begin{equation}
\mathcal{E}_{+}(t_{c}) = \mathcal{E}_{-}(t_{c}).
\label{eqn:complexzeroeq}
\end{equation}
The method of Davis and Pechukas was to move away from the real time axis to the upper half of the complex plane and integrate the Schr{\"o}dinger equation in the adiabatic basis along the level line of  $D(t)$ which is defined by $\mathfrak{Im} \left[ D(t) \right] = \mathfrak{Im} \left[ D(t_{c})\right]$. The main assumptions allowing this approach were that there is no crossings for real $t$, that $t_{c}$ is well separated from other zero points or possible singularities and that the Hamiltonian is analytic and single-valued at least in a region of complex $t$ plane bounded by the real axis and the level line mentioned above. 
It is evident from Eq.~(\ref{eqn:adiabaticlevels}) that the points $t_{c}$ satisfying (\ref{eqn:complexzeroeq}) coincide with the zeros of the adiabatic levels. As discussed earlier, crossings of the adiabatic levels are typically avoided ones when $t$ is a real variable, so $t_{c}$ is in general complex and usually a branch point of the eigenenergy. 

In the case that there are many zero points $t_{c}$, Eq. (\ref{eqn:dykhne}) has to be complemented accordingly and there exists some rigorous results on the matter \cite{Joye1991, Joye1991_2}. However, as discussed already in the seminal paper of Davis and Pechukas \cite{dp} and studied later by Suominen and co-workers \cite{kasphd, vitanov, artikkeli}, including the contribution of all the zero points on the half plane as a coherent sum can be very useful in order to obtain a good approximation for $P$ when the system parameters are outside the adiabatic region. Therefore, we define the generalization of the DDP formula as
\begin{equation}
P_{DDP} = \left\vert \sum_{k = 1}^{N} \Gamma_{k} e^{i D\left(t_{c}^{k}\right)} \right\vert^{2}, 
\label{eqn:genDDP}
\end{equation}
where
\begin{equation}
\Gamma_{k} = 4 i \lim_{t \rightarrow t_{c}^{k}} \left( t - t_{c}^{k}\right) \gamma(t),
\label{eqn:isogamma}
\end{equation}
and $\gamma(t)$ is the nonadiabatic coupling defined in Eq.~(\ref{eqn:nonadiabcoupl}). In particular, it was found in \cite{artikkeli} that this definition is necessary to approximate the oscillatory behavior of the final transition probability of the superparabolic level-glancing models. Indeed, from Eq.~(\ref{eqn:genDDP}) one can be see that the existence of multiple zero points $t_{c}^{k}, \, k = 1, 2, 3, \ldots$ leads to oscillations in the final state populations as the parameters are varied. 

\subsubsection{Application to superparabolic models}

For the case of superparabolic level-glancing model, the generalized DDP formula is obtained and discussed in detail in the reference \cite{artikkeli} and are simply given below. The zero points of the eigenvalues of the Hamiltonian defined by Eq.~(\ref{eqn:SPLGcmodel}) are
\begin{equation}
t_{c}^{k} = \alpha^{1/N} e^{i \pi (2 k - 1)/ 2N}, \; k = 1, 2, \ldots, N,
\end{equation}
so the zero points lie on a circle of radius $\alpha^{1/N}$ in the complex t-plane. The integrals over the adiabatic energies are given by
\begin{equation}
D(t_{c}^{k}) = \eta e^{i \pi (2k - 1)/2N},
\label{eqn:Dcalculated}
\end{equation}
where
\begin{equation}
\eta = 2 \nu_{N}\alpha^{(N + 1)/N}
\end{equation}
and
\begin{equation}
\nu_{N} = \int_{0}^{1} \sqrt{1 - y^{2N}} dy = \frac{1}{2N} B\left( \frac{1}{2N}, \frac{3}{2}\right),
\end{equation}
where $B(x, y)$ is the $\beta$ function~\cite{NIST}. These explicit expressions are also used when discussing ZNT as they form the time-dependent version of the phase integrals in that context. 
The factors in Eq.~(\ref{eqn:isogamma}) are given by $ \Gamma_{k} = (-1)^k$ and the points $t_{c}^{k}$ are grouped into pairs with the same imaginary part  to give the generalized DDP formula in the form
\begin{widetext}
\begin{equation}
P_{DDP} = 4 \left\vert \Sigma_{k=1}^{N/2} (-1)^{k} e^{- \eta \sin \left[\frac{\pi}{2N}(2k - 1)\right]} \sin \left[\eta \cos{\frac{\pi}{2N} (2k - 1)} \right] \right\vert^{2}.
\label{eqn:ddpsuperparabolinen}
\end{equation}
\end{widetext}

\section{Zhu-Nakamura Theory}
\label{sec:znt}
\subsection{Backround and exact results for the LZ case}

Determining the probability of nonadiabatic transitions can be reduced to calculating the Stokes constants as mentioned in the introduction. This way, one obtains the connection between the different fundamental asymptotic solutions of the differential equation governing the process, that are valid in different regions of the complex plane \cite{FedoryukBook}. From these constants, one can construct the scattering matrix $\left(S\left( a, b \right)\right)_{mn}$ (usually called  transition matrix in time-dependent problems) for the process. The transition probability is given in terms of the reduced scattering matrix elements $\left(S^{R}\right)_{mn}$ as $P_{12} := \vert \left(S^{R}\right)_{12} \vert^{2}$. In practice, the Stokes constants are known only for very restricted class of differential equations.

In Ref.~\cite{ZNStokesConstants}, Zhu and Nakamura calculated the Stokes constants for four different classes of second-order differential equations. Among them was the important special cases of
\begin{equation}
y''(z) +q(z)y(z) = 0,
\end{equation}
where $q(z)$ is either a quartic polynomial or a polynomial where the degree of the highest term is $2 n $ and the next highest term is of degree $n - 1$, where $n$ is a positive integer. This means that the solved cases contained both the time-independent LZ model and the time-dependent parabolic model as well as the superparabolic level-glancing models.

The  important general results of this work were that the Stokes constants $U_{i}, \quad i = 1, ... , 2(n + 1) $ could be expressed in terms of only one of them, say $U_{1}$, and that this $U_{1}$ could be expressed as a converging infinite series depending on the constants of the polynomial $q(z)$. Unfortunately, the expression for $U_{1}$ is too cumbersome and so the results of \cite{ZNStokesConstants} are not very transparent to analytical analysis and therefore of only limited practical value (See, e.g., the discussions of their method in \cite{Osherov2010} and \cite{CrothersBook}). 
However, from the exact results one could also obtain the general form of the scattering matrix elements in terms of the Stokes constant $U_{1}$ so that for example for the LZ type problem (see the subsection \ref{subsec:time-independent}) we have the relation 
\begin{equation}
\left( S_{LZ}^{R} \right)_{12} = -\frac{2\dot{\imath} {\rm Im}\left(U_{1}\right)}{1 + \vert U_{1} \vert^{2} }.
\end{equation}  
Furthermore, when the transition probability for one passage of the crossing point is denoted as
\begin{equation}
p = \frac{1}{1 +\vert U_{1} \vert^{2} }
\end{equation}
we get the exact result concerning the functional form of the overall transition probability as
\begin{equation}
P_{12} = 4 p (1 - p) \sin^{2} (\psi),
\label{eqn:functionalform}
\end{equation}
where $\psi = arg \left( U_{1}\right)$. This form is of course similar to the equation for double-passage transition probability derived by St{\"u}ckelberg in the study of atomic collisions \cite{Stuckelberg1932}.

In order to overcome the fact that the formulas for the Stokes constants are too difficult to be practical, Zhu and Nakamura have considered semiclassical approximations together with "ad hoc" experimental modifications to obtain final formulas for the probability of nonadiabatic transitions applicable to general  situations. Thus the content of ZNT is, in a way, reduced to a set of relatively simple and compact approximate formulas that are obtained by first taking the exact functional form (such as Eq.~(\ref{eqn:functionalform})) as the starting point and then making any modification to its constitutive parts in order to obtain a good approximation. The motivation and guideline behind the construction of the approximate formulas was not only to formulate a general theory that would overcome the deficiencies of the previously existing theories of nonadiabatic transitions but also to give the final formulas in a simple and compact form that would avoid altogether the complex calculus and contour integration and only in terms of the adiabatic potentials. This has of course relevance in particular to experimentalists and in multilevel situations \cite{tdznt, znPhaseInt}.  

\subsection{The final recommended formulas of ZNT}

For the direct application of the ZNT, Zhu and Nakamura offer the final formulas to be used in a general level-crossing situation, whether time-independent or time-dependent. These have appeared in different forms over the years, but we take the definitive ones to be those given in the appendices of \cite{NakamuraBook}. We consider here first the parabolic level-glancing case and then discuss applicability of the general formulas for the superparabolic models. 
As the parabolic model is the one also studied explicitly by Nakamura and co-workers, we expect the formulas work well in this first case. Their viewpoint at the time was however different as they considered time-independent processes and the transition probability for a variable collision energy and fixed values of the coupling strength. On the contrary, we have effectively fixed $b^{2}$ ($b = 0$) and $a^{2}$ is related through Eq.~(\ref{eqn:relparam}) to the independent variable in our case. Also, using that relation, it is seen that the parameter range covered here is much larger. The superparabolic models are very similar in character to the parabolic ones but need to be dealt within the context of ZNT by the formulas meant for general models, so it is interesting to see how well this transition works. 

The final formulas for the time-independent LZ case is given for two parameter values separately, when energy is higher than the crossing energy ($E \geq E_{X}$) or lower than the crossing energy ($E \leq E_{X}$). As this corresponds to either non-negative or non-positive $b^{2}$ (i.e., the cases include equality)
, we could in principle use either one of the final formulas. The formulas for $E \geq E_{X}$ are somewhat more compact and considered here first. 

\subsubsection{Double-crossing formulas for the parabolic level-glancing model}

The overall probability of nonadiabatic transition is now still given by (\ref{eqn:functionalform}) but the different terms are given by the following expressions:
The modification of the Landau-Zener formula is given by
\begin{equation}
p = \exp\left[ -\frac{\pi}{4a}\left(\frac{2}{b^{2} + \sqrt{b^{4} + 0.4a^{2} + 0.7}}\right)^{1/2}\right],
\label{eqn:modlz}
\end{equation}
and the phase is given by
\begin{equation}
\psi = \sigma + \phi_{s} = \sigma - \frac{\delta}{\pi} + \frac{\delta}{\pi}\ln\left(\frac{\delta}{\pi}\right) - arg \Gamma\left(\dot{\imath}\frac{\delta}{\pi}\right) - \frac{\pi}{4}.
\label{eqn:psilz}
\end{equation}
This contains the real and imaginary parts of the phase integral
\begin{equation}
D(t_{c}^{1}) = \sigma + \dot{\imath}\delta.
\label{eqn:phaseinttd}
\end{equation}
Furthermore, it is advantageous to replace this imaginary part $\delta$ in (\ref{eqn:psilz}) by the modification
\begin{equation}
\delta_{\psi} = \left( 1 + \frac{5\sqrt{a}}{\sqrt{a} + 0.8}10^{-\sigma}\right)\delta .
\label{eqn:moddelta}
\end{equation}

This modifies the phase (\ref{eqn:psilz}) for intermediate and large values of $a$ or, equivalently, for small and intermediate values of $\alpha$.
Note how the probability of a nonadiabatic transition for one passage of the crossing point (\ref{eqn:modlz}) differs from the original LZ formula (\ref{eqn:lz_formula}). It follows, that the ZNT could in principle work also in the case $b = 0$. 
It should also be noted that the form of Eqs.~(\ref{eqn:modlz}) and (\ref{eqn:moddelta}) are obtained completely heuristically, i.e., it is not derived from anywhere (see, e.g., Ref.~\cite{zn101}). The Zhu-Nakamura theory actually also offers further approximations to avoid the complex integration in (\ref{eqn:phaseinttd}). However, because the phase integral (\ref{eqn:phaseinttd}) appears also in the DDP formulas it was calculated exactly for any complex crossing point of any superparabolic level-glancing model in \cite{artikkeli} and it is given by (\ref{eqn:Dcalculated}), so we do not need those approximations here but will discuss about them below along with other modifications when considering the superparabolic models. Let us now use these results for the parabolic level glancing model.

The transition probability for one passage is now given as
\begin{equation}
p = \exp\left[ -\frac{\pi \alpha^{3/2}}{\sqrt{2}}\left(\frac{1}{(0.1\alpha^{-3} + 0.7)^{1/4}}\right)\right].
\end{equation}

For the parabolic level-glancing model, the real and imaginary parts of the phase integral are equal as can be seen from Eq.~(\ref{eqn:Dcalculated}) and we simply have
\begin{equation}
\sigma = \delta = c \alpha^{3/2},
\label{eqn:deltaLG}
\end{equation}
where we have defined the constant
\begin{equation}
c =  \frac{\sqrt{\pi} \Gamma(1/4)}{3\sqrt{2} \Gamma(3/4)}.
\end{equation}
The final transition probability can now be explicitly stated
\begin{widetext}
\begin{equation}
\begin{aligned}
P_{12} &= 4 e^{-\frac{\pi \alpha^{3/2}}{\sqrt{2}}(0.1\alpha^{-3} + 0.7)^{-1/4}} \left( 1 - e^{-\frac{\pi \alpha^{3/2}}{\sqrt{2}}(0.1\alpha^{-3} + 0.7)^{-1/4}} \right)\times \\ 
& \times\sin^{2}\left[ c\alpha^{3/2} - \frac{ c\alpha^{3/2}}{\pi} + \frac{ c\alpha^{3/2}}{\pi}\ln\left(\frac{ c\alpha^{3/2}}{\pi}\right) - arg \Gamma\left(\dot{\imath}\frac{ c\alpha^{3/2}}{\pi}\right) - \frac{\pi}{4}\right]
\end{aligned}
\label{eqn:znresult}
\end{equation}
\end{widetext}
This can be compared to the DDP result for the parabolic glancing model, which now explicitly reads
\begin{equation}
P_{DDP} = 4 e^{-2 c \alpha^{3/2}}\sin^{2} \left[ c \alpha^{3/2}\right].
\label{eqn:ddpresult}
\end{equation}

We can compare the forms of the Eqs.~(\ref{eqn:ddpresult}) and (\ref{eqn:znresult}) and one could think that the argument in the sine function of the DDP result is just the first term of the corresponding term in the ZNT result. Furthermore, considering that the exact total probability is of the form $P = 4p(1-p)\sin^{2}(\psi)$ one could guess that the exponential in DDP result could be interpreted as $p$. However, although its behaviour is similar to the exponential in ZNT result, modifying DDP result this way does not offer an improvement to the approximation.

\subsubsection{Tunneling formulas for the parabolic level-glancing model}
Again, the general formula 
\begin{equation}
P_{12} = 4 p (1 - p)\sin^{2} (argU_{1})
\end{equation}
applies, but now with
\begin{equation}
p = \frac{1}{1 + B(\sigma / \pi)e^{2\sigma} - g_{2}\sin^{2}(\sigma)}
\end{equation}
and the Stokes constant as
\begin{equation}
\textrm{Re}(U_{1}) = \cos (\sigma)\left[ \sqrt{B(\sigma / \pi)}e^{\sigma} - g_{1}\sin^{2}(\sigma)\frac{e^{-\sigma}}{\sqrt{B(\sigma / \pi)}}\right]
\end{equation}
and
\begin{widetext}
\begin{equation}
\textrm{Im}(U_{1}) = \sin (\sigma)\left[ B(\sigma / \pi)e^{2\sigma} - g_{1}^{2}\sin^{2}(\sigma) \cos^{2}(\sigma)\frac{e^{-2\sigma}}{B(\sigma / \pi)} + 2 g_{1}\cos^{2}(\sigma) - g_{2} \right]^{1/2},
\end{equation}
\end{widetext}
and where the function $B(x)$ is not to be confused with $\beta$ function but is defined as
\begin{equation}
B(x) = \frac{2\pi x^{2x}}{x \Gamma^{2}(x)}.
\end{equation}
The functions $g_{1}$ and $g_{2}$ are again ad hoc modifications and are given by 
\begin{equation}
g_{1} = 1.8\left(a^{2}\right)^{0.23}\exp\left(-\delta \right),
\end{equation}
and
\begin{equation}
g_{2} = \frac{3\sigma}{\pi \delta}\ln \left(1.2 + a^{2}\right) - \frac{1}{a^{2}},
\end{equation}
and were both actually equal to unity in the original time-independent LZ case. The constant $b^{2}$ does not appear in these formulas explicitly but only through the phase integrals $\delta$ and $\sigma$. The final formula for the parabolic level-glancing model in this tunneling case is therefore obtained just by substituting those terms obtained from Eqs.~(\ref{eqn:deltaLG}) and (\ref{eqn:relparam}) into the above formulas.  

\subsubsection{Application to superparabolic models}

The above expressions dealt only with quadratic time-dependencies, but ZNT can be applied to models with general time-dependencies by using these same formulas but with replacing the phase integral terms $\sigma$ and $\delta$ with the phase integral of each model \cite{tdznt}. 
Also, the diabatic parameters (in our case, only the diabatic coupling $\alpha$) should be modified, either by fitting the  potential to a parabolic one or by using directly the final formulas given in \cite{NakamuraBook}:
\begin{equation}
a^{2} = \frac{\sqrt{d^{2}-1}\hbar^{2}}{2V_{0}^{2}\left(t_{t}^{2} - t_{b}^{2}\right)}, \quad b^{2} = \sqrt{d^{2}-1} \frac{t_{t}^{2} + t_{b}^{2}}{t_{t}^{2} - t_{b}^{2}},
\label{eqn:paramfit1}
\end{equation}
and the quantities in the formulas are given as
\begin{equation}
V_{0} = (E_{2}(t_{0}) - E_{1}(t_{0}))/2,
\label{eqn:paramfit2}
\end{equation}
and
\begin{equation}
d^{2} = \frac{\left[ E_{2}(t_{b}) - E_{1}(t_{b})\right]\left[ E_{2}(t_{t}) - E_{1}(t_{t})\right]}{\left[ E_{2}(t_{0}) - E_{1}(t_{0})\right]^{2}},
\label{eqn:paramfit3}
\end{equation}
where $E_{2}(t) > E_{1}(t)$ are the adiabatic potentials, $t_{b}$ is the moment when $E_{2}$ reaches minimum, $t_{t}$ when $E_{1}$ reaches maximum and $t_{0}$ when the difference between the adiabatic energies is minimum. The philosophy behind this in the ZNT is that the formulas (\ref{eqn:paramfit1})-(\ref{eqn:paramfit2}) only refer to adiabatic quantities so one could take the experimentally measured adiabatic energies $E_{i}$ and 
fit them. We can of course use the adiabatic model parameters given by Eqs.~(\ref{eqn:adiabaticlevels}) and (\ref{eqn:SPLGcmodel}).

In the case of superparabolic level glancing models, $t_{b, t, 0} = 0$, so these several points actually coincide and $d^{2}$ becomes unity. It is then clear from the above relations that the direct application of the final formulas of ZNT is not possible because of the several ambiguous "0/0" type relations.  The diabatic coupling (\ref{eqn:paramfit2}), however,  gives $V_{0} = \alpha$ as it should.

The phase integral for the superparabolic models was calculated in (\ref{eqn:Dcalculated}). However, ZNT also offers an approximate expression for its calculation in its collection of final recommended formulas in \cite{NakamuraBook} as
\begin{equation}
\sigma + \dot{\imath}\delta = \frac{1}{\hbar}\left[ \int_{0}^{t_{b}} E_{2}(t)dt - \int_{0}^{t_{t}} E_{1}(t)dt +\sqrt{\frac{b^{2}}{a^{2}}}+ \Delta \right]
\end{equation}
where
\begin{align}
\Delta =& \frac{t_{0} -(t_{b} + t_{t})/2}{\sqrt{a^{2}(b^{4} + \dot{\imath}})(t_{b} - t_{t})}\sqrt{\frac{d^{2}}{d^{2} - 1}} \nonumber \\ &+ \frac{1}{2\sqrt{a^{2}}}\int_{0}^{\dot{\imath}}\left( \frac{1 + t^{2}}{t + b^{2}}\right)^{1/2} dt,
\end{align}
and it is obvious that the ambiguities arise also here. 

As also the first strategy, namely fitting the superparabolic diabatic energy level to a parabolic one, is doomed to fail (for a similar reason why the model cannot be linearized at the glancing and dealt with by the time-dependent LZ model \cite{Suominen1992}), and as it is stated in \cite{tdznt} that once $\sigma$ and $\delta$ are obtained the final results are not so sensitive to other dependencies of the parameters. Motivated by this, we take the ZNT formulas of the previous subsections with the formula (\ref{eqn:Dcalculated}) to study how well ZNT generalizes to the superparabolic models in the next section. 

\section{Results}
\label{sec:results}

\begin{figure}
\begin{center} 
 \includegraphics[scale=0.65]{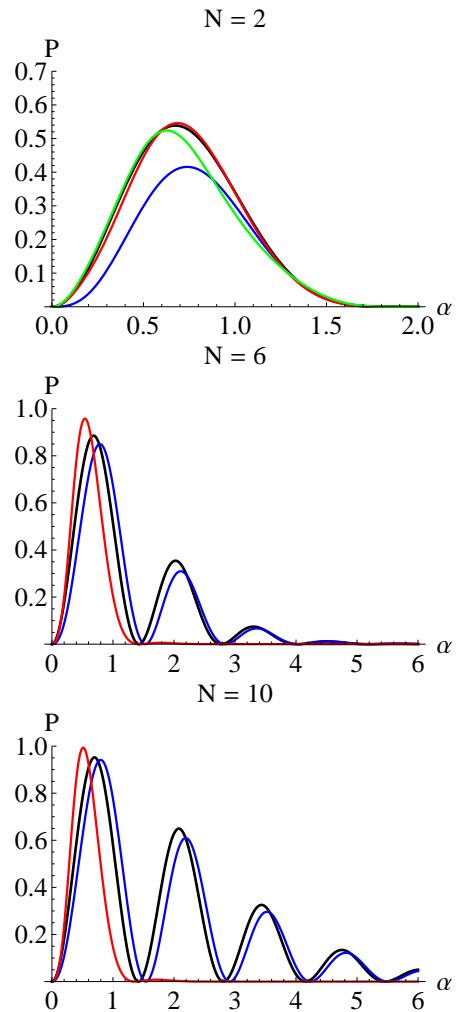} 
\caption{(Color online) The DDP approximation is the blue line, ZNT (diabatic crossing case) is the red and the numerical result is the black line. The green line in the first plot is the ZNT formula for tunneling case. It does not generalize well at all for higher values of $N$ and is omitted.}
\label{fig:diabaticZN}
\end{center}
\end{figure}
\begin{figure}
\begin{center} 
 \includegraphics[scale=0.68]{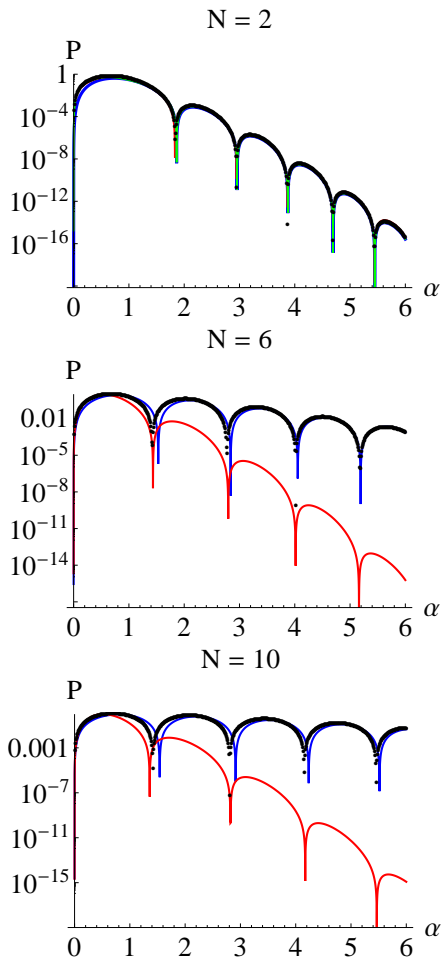} 
\caption{(Color online) This is similar to the previous figure but with logarithmic ordinate.}
\label{fig:diabaticZNlog}
\end{center}
\end{figure}
The results of the DDP theory and the ZNT are compared to the results of the numerical simulations in Figs.~\ref{fig:diabaticZN} and \ref{fig:diabaticZNlog}. We take the example cases to be the parabolic level-glancing model and the superparabolic level-glancing models with $N = 6$ and $N = 10$. 

As expected, the parabolic model is well approximated by the ZNT theory, particularly by the double-crossing formula which overlaps with the numerical result. The approximate formula derived for the tunneling type transitions in ZNT as well as the DDP formula give the right characteristic features everywhere and tend to the exact result in the adiabatic region. It is also interesting to note that the two approximate formulas of the ZNT do not coincide.

When going to the higher values of $N$, it is seen that the ZNT does not give a good approximation. The tunneling formulas are omitted altogether from the middle and bottom plots in Figs.~\ref{fig:diabaticZN} and \ref{fig:diabaticZNlog} as they give nonsensical behavior. For the double-crossing formulas, the exponentially decaying part dies of too quickly and there is only one peak visible in the linear transition probability plot in Fig.~\ref{fig:diabaticZN}. Interestingly however, it is seen from the logarithmic plot in Fig.~\ref{fig:diabaticZNlog} that the oscillations of the transition probability seem to have the right frequency, so the Stokes phase of the LZ model with the modifications of the ZNT seems to work well and the problem is in the expression for the probability of one transition, i.e., in Eq.~(\ref{eqn:modlz}). 

The DDP formula on the other hand generalizes easily for the higher superparabolic models (i.e. higher values of $N$) and apart from the small phase shift is a very good approximation in the whole parameter region. 

\section{Conclusions}
\label{sec:conclusions}

We have reviewed the basic aspects of ZNT and studied their application to the superparabolic models. Although ZNT has been studied in the context of the parabolic model before, we considered this case from a different viewpoint than done in the original works of Nakamura, Zhu and co-workers. Our treatment was explicitly time-dependent and we studied the transition probability as a function of the diabatic coupling, instead of the usual treatment of the equivalent collisional model where the independent variable is the collision energy and the coupling is held fixed. Of course, these are just two opposite viewpoints, so in effect we also studied the whole zero-energy parameter region for a wider range of values for the coupling. Obtaining approximative expressions for this region was of course one of the major motivating factors behind the whole Zhu-Nakamura approach, as it was not properly dealt with by the previously-existing theories. 

The level-glancing case resembles in many ways the double-crossing one, as there are oscillations in the final populations and the adiabatic coupling has two distinct peaks, for example. Indeed, the ZNT approximation derived for the double-crossing case is very accurate in the parabolic level-glancing model. As the above-mentioned behavior is common to all of the superparabolic models it is somewhat surprising that the ZNT at its present does not work well for the models with higher values of $N$ but that at least the expression for $p $ (Eq.~(\ref{eqn:modlz})) should be modified further. It also seems clear, that the compact formulation of the ZNT for general models, namely, the parameter-fitting formulas (\ref{eqn:paramfit1}) - (\ref{eqn:paramfit3}), do not seem to be straightforwardly applicable to adiabatic energies that are symmetric in time, i.e., when $t_{0}$, $t_{b}$ and $t_{t}$ coincide.

At the same time one can see that the generalization of the DDP theory, namely, the full summation formula (\ref{eqn:genDDP}), can be directly applied in a systematic fashion and it does not lead to difficulties in the conceptual level. Of course it also lacks of a sufficient general mathematical proof that would include superparabolic models and more work remains to be done in this direction. 
On the other hand, it may be more difficult to handle multilevel problems within DDP theory \cite{znPhaseInt}. Furthermore, DDP theory is not so directly related quantities measured in experiments as ZNT due to the fact that it relies on analytic continuation of the adiabatic energies.     
Therefore, it has been our point to show the usefulness of the both of these approaches and especially to highlight the need for more general formulation of the ZNT and also the need for more careful instructions to its application.



\begin{acknowledgments}
This research was supported by the Finnish Academy of Science and Letters, and the Academy of Finland, grant 133682. The author would like to thank K.-A.  Suominen for discussions.
\end{acknowledgments}


\begin{thebibliography}{50}
\bibitem{NakamuraBook} H. Nakamura, \textit{Nonadiabatic Transition: Concepts, Basic Theories and Applications, 2nd ed.}, (World Scientific, Singapore, 2012).
\bibitem{Child1974} M. S. Child, \textit{Molecular Collision Theory}, (Academic Press, London, 1974).
\bibitem{kasphd} K.-A. Suominen, Ph.D. thesis, University of Helsinki, Finland, 1992.
\bibitem{Zener1932} C. Zener, Proc. R. Soc. London, A \textbf{137}, 696 (1932).
\bibitem{Landau1932} L. D. Landau, Phys. Z. Sowjet Union \textbf{2}, 46 (1932).
\bibitem{Stuckelberg1932} E. C. G. St{\"u}ckelberg, Helv. Phys. Acta {\bf 5}, 369 (1932).
\bibitem{Majorana1932} E. Majorana, Nuovo Cimento \textbf{9}, 43 (1932); G. F. Bassani (ed.), \textit{Ettore Majorana: Scientific Papers}, (SIF, Bologna, 2006).
\bibitem{kazantsev} A. P. Kazantsev, G. I. Surdutovich and V. P. Yakovlev, \textit{Mechanical Action of Light on Atoms}, (World Scientific, Singapore, 1990).
\bibitem{Gaitan2003} F. Gaitan, Phys. Rev. A \textbf{68}, 052314 (2003).
\bibitem{Dziarmaga2010} J. Dziarmaga, Adv. Phys. \textbf{59}, 1063 (2010).
\bibitem{Polkovnikov2011} A. Polkovnikov, K. Sengupta, A. Silva, and M. Vengalattore, Rev. Mod. Phys. \textbf{83}, 863 (2011).
\bibitem{ZNTeranishi2001} C. Zhu, Y. Teranishi and H. Nakamura, Adv. Chem. Phys. \textbf{117}, 127 (2001).
\bibitem{ZNStokesConstants} C. Zhu and H. Nakamura, J. Math. Phys. \textbf{33}, 2697 (1992).
\bibitem{Hinton1979} F. L. Hinton, J. Math. Phys. {\bf 20}, 2036 (1979).
\bibitem{Zou2005} S. Zou, A. Kondorskiy, G. Mil'nikov, and H. Nakamura, J. Chem. Phys. \textbf{122}, 084112 (2005).
\bibitem{Chang2010} Bo Y. Chang, S. Shin, and I. R. Sola, Phys. Rev. A \textbf{82}, 063414 (2010); Bo Y. Chang, S. Shin, J. Santamaria, and I. R. Sola, J. Phys. Chem. A \textbf{116}, 2691 (2012).
\bibitem{tdznt} Y. Teranishi and H. Nakamura, J. Chem. Phys. \textbf{107}, 1904 (1997).
\bibitem{Crothers1977} D. S. F. Crothers and J. G. Hughes, J. Phys. B: Atom. Molec. Phys. \textbf{10}, L557 (1977).
\bibitem{Suominen1992} K.-A. Suominen, Optics Comm. \textbf{93}, 126 (1992).
\bibitem{Shimshoni1991} E. Shimshoni and Y. Gefen, Ann. Phys. \textbf{210}, 16 (1991).
\bibitem{Ronveaux} A. Ronveaux, \textit{Heun's Differential Equation}, (Oxford University Press, New York, 1995).
\bibitem{ZNNobusada2001} C. Zhu, K. Nobusada and H. Nakamura, J. Chem. Phys. \textbf{115}, 3031 (2001).
\bibitem{Nakamura2008} H. Nakamura, Adv. Chem. Phys. \textbf{138}, 95 (2008).
\bibitem{artikkeli} J. Lehto and K.-A. Suominen, Phys. Rev. A \textbf{86}, 033415 (2012).
\bibitem{NikitinReview} E. E. Nikitin, Annu. Rev. Phys. Chem. \textbf{50}, 1 (1999).
\bibitem{Fuchs2012} J.-N. Fuchs, L.-K. Lim and G. Montambaux, Phys. Rev. A \textbf{86}, 063613 (2012).
\bibitem{zn101}  C. Zhu and H. Nakamura, J. Chem. Phys. \textbf{101}, 4855 (1994).
\bibitem{dp} J. P. Davis and P. Pechukas, J. Chem. Phys. \textbf{64}, 3129 (1976).
\bibitem{dykhne} A. M. Dykhne, Sov. Phys. JETP \textbf{11}, 411 (1960); \textbf{14}, 941 (1962).
\bibitem{Joye1991} A. Joye, H. Kunz and Ch.-Ed. Pfister, Ann. Phys. \textbf{208}, 299 (1991).
\bibitem{Joye1991_2} A. Joye, G. Mileti and Ch.-Ed. Pfister, Phys. Rev. A \textbf{44}, 4280 (1991). 
\bibitem{vitanov} N. V. Vitanov and K.-A. Suominen, Phys. Rev. A \textbf{59}, 4580 (1999).
\bibitem{NIST} F. W. J. Olver, D. W. Lozier, R. F. Boisvert, and C. W. Clark, \textit{NIST Handbook of Mathematical Functions}, (Cambridge University Press, Cambridge, 2010).
\bibitem{FedoryukBook} M. V. Fedoryuk, {\it Asymptotic Analysis: Linear Ordinary Differential Equations}, (Springer-Verlag, Heidelberg, 1993).
\bibitem{Osherov2010} V. I. Osherov and V. G. Ushakov, J. Phys. A: Math. Theor. \textbf{43}, 145203 (2010).
\bibitem{CrothersBook} D. S. F. Crothers, \textit{Semiclassical Dynamics and Relaxation}, (Springer, New York, 2008).
\bibitem{znPhaseInt} C. Zhu and H. Nakamura, J. Chem. Phys. \textbf{109}, 4689 (1998).



\end{thebibliography}
\end{document}